\documentclass[journal,comsoc]{IEEEtran}
\IEEEoverridecommandlockouts
\usepackage{amssymb}
\usepackage{amsmath}
\usepackage{caption}
\usepackage{cite}
\usepackage{mathrsfs}
\usepackage[displaymath,mathlines]{lineno}
\usepackage{color}
\usepackage{overpic}
\usepackage{algorithm,algpseudocode}
\usepackage{algorithmicx}
\usepackage{pbox}
\usepackage{multicol}
\usepackage{amsthm}
\usepackage{epstopdf}
\usepackage{xcolor}
\usepackage{ulem}
\makeatletter
\newcommand{\doublewidetilde}[1]{{%
		\mathpalette\double@widetilde{#1}%
}}
\newcommand{\double@widetilde}[2]{%
	\sbox\z@{$\m@th#1\widetilde{#2}$}%
	\ht\z@=.5\ht\z@
	\widetilde{\box\z@}%
}
\makeatother

\definecolor{red}{rgb}{0,0,0}
\begin{document}
\title{Digital Twin for O-RAN Towards 6G}
\author{Huan X. Nguyen, Kexuan Sun, Duc To, Quoc-Tuan Vien, and Tuan Anh Le 
\thanks{H. X. Nguyen, Q.-T. Vien, and T. A. Le are with the London Digital Twin Research Centre and 5G/6G Research Group, Middlesex University, London, NW4 4BT, U. K. Email: \{h.nguyen; q.vien; t.le\}@mdx.ac.uk. H. X. Nguyen is also with International School, Vietnam National University, Hanoi, Vietnam.}
\thanks{K. Sun and D. To are with Rakuten Symphony UK, the Height, Weighbridge, Surrey, KT13 ONY, UK. Email : \{kexuan.sun; duc.to\}@rakuten.com.}
\thanks{This work is supported by the EPSRC UK-India Future Networks Initiative (Grant ref-EP/W016524/1)}
}

\markboth{IEEE  Communications Magazine, DOI: 10.1109/MCOM.003.2400016 }
{Shell \MakeLowercase{\textit{et al.}}: Bare Demo of IEEEtran.cls for IEEE Journals}

\maketitle

\begin{abstract}

In future wireless systems of beyond 5G and 6G, addressing diverse applications with varying quality requirements is essential. Open Radio Access Network (O-RAN) architectures offer the potential for dynamic resource adaptation based on traffic demands. However, achieving real-time resource orchestration remains a challenge. Simultaneously, Digital Twin (DT) technology holds promise for testing and analysing complex systems, offering a unique platform for addressing dynamic operation and automation in O-RAN architectures. Yet, developing DTs for complex 5G/6G networks poses challenges, including data exchanges, ML model training data availability, network dynamics, processing power limitations, interdisciplinary collaboration needs, and a lack of standardized methodologies. This paper provides an overview of Open RAN architecture, trend and challenges, proposing the DT concepts for O-RAN with solution examples showcasing its integration into the framework.
\end{abstract}
%\begin{IEEEkeywords}
%Digital Twin,  O-RAN, 6G, DT-RAN, RIC
%\end{IEEEkeywords}
\IEEEpeerreviewmaketitle

\section{Introduction}
At the heart of digital transformation lies the concept of the Digital Twin (DT), a virtual replica of a physical system \cite{Mihai_et_al:2022:DTS}. Throughout its life cycle, the physical system and its DT continuously exchange data. This ability to simulate and analyse the physical system's behaviours in real-time via its DT enables testing and optimization without risking interference with the corresponding physical asset. In wireless communications, the role of DTs can be crucial \cite{NA23}, particularly in facilitating Open Radio Access Network (O-RAN) to deliver high-quality services to end users \cite{Nguyen_et_al:2021:DT5G,MY23}.

Towards 2030, the next generation of radio system, known as 6G, will be evolved to support unprecedented scenarios including integrated sensing and communications, integrated AI, and ubiquitous connectivity. The increasing complexity of networks necessitates an acceleration in the adoption of open architectures for existing RANs \cite{ORAN_Alliance:2018:ORAN}. In response, DT technology emerges as a crucial enabler in O-RAN architecture, facilitating better understanding, optimization, and management of network elements, ultimately contributing to the enhancement of 6G systems 
%\cite{Report:2022:ITU_R_M_2516_0,
\cite{Lin_et_al:2023:6GDT,Masaracchia_et_al:2023:DTO,Zhenyu2023}.\footnote{Wider application range of DT on Open RAN can potentially include other use cases such as real-time traffic management and steering, adaptive energy efficiency, intelligent resource management in dynamically sliced networks \cite{MY23}, and radio network coverage planning and optimisation \cite{Lin_et_al:2023:6GDT}. %,Zhengming_2023}. 
}

Within an O-RAN network, DT technologies enable the generation of a virtual model replicating part of or the entire RAN infrastructure, covering logical functions of base stations, hardware components, and end-user devices. This virtual representation facilitates simulation and analysis of diverse network scenarios, including traffic patterns, congestion, and interference. The integration of the DT concept into O-RAN aligns perfectly with the principles of openness, autonomy, and intelligence that are foundational to O-RAN. It enables network operators to simulate, analyse, and optimise network behaviour in real-time. As 5G and 6G networks continue to expand rapidly, the DT is set to become increasingly vital in the wireless communications industry, playing a key role in delivering innovative and reliable next-generation services to end-users.

While the literature extensively defines the concept of DT, its integration into O-RAN remains currently under early development. Example includes the work by Mirzaei et. al.  \cite{Mirzaei2023network} which outlined potential benefits of a network of DT, illustrated by two practical use cases of energy saving and traffic steering, in both the planning and operation phases of a RAN. There is an initial prototype in the Colosseum project \cite{Polese2024}, that functions the RAN elements and various RF channel conditions as a twin for a real network. The testbed offers the capability of simulating the transmissions between nodes with 4G/5G protocol stacks over a RF digital emulator to generate data for AI/ML applications or interact with other radio testbeds. In our work, we provide further insights into how a DT network can be built and integrated into the RAN with open architecture. Especially, this article addresses the formulation of DT either as a virtual replica of or as building blocks within the RAN, highlighting both potentials and research challenges associated with its integration into the O-RAN framework. We will propose and discuss relevant DT concepts within the O-RAN architecture, presenting solution examples and outlining future research directions. The paper aims to establish the groundwork for implementing DT concepts, supporting the automation, optimization, and enhancing the operational efficiency of O-RAN for next-generation wireless networks.

\section{O-RAN: Trend and Architecture Towards Intelligent Control}

\subsection{Trend of Disaggregation and Openness for RAN}

The RAN consists of a network of base stations (BSs), each covering a geometrical cell to provide connectivity for user equipments (UEs) within that area. Together with the core network (CN), it establishes connections, interfaces with other networks, and facilitates wireless transmissions for mobile users. Traditionally, a BS, usually supplied as a complete package by a single vendor, converts digital signals from the CN into high-frequency analog signals for over-the-air downlink transmission (and a reversed process for the uplink). However, recent BS designs have evolved by separating the units responsible for signal conversion from those handling digital processing. This allows for more compact and easily installable radio units near the antenna towers. It also facilitates virtualisation, enabling the sharing of digital processing tasks among multiple BSs on a single computational pooling-based platform through edge cloud computing.

This trend has given rise to the O-RAN movement, which exemplifies a significant shift towards openness and interoperability within the telecommunications industry. It redefines the architectural components of a BS and standardises the interfaces between these components, allowing software-driven functionalities to be more easily integrated and managed, independent of the underlying hardware platforms. This openness not only promotes a more competitive and diverse market but also encourages innovation by allowing a multitude of vendors to contribute and compete, breaking  the traditional vendor lock-ins, lowering entry barriers for new players, and promoting a vibrant ecosystem of interoperable solutions that can adapt swiftly to evolving market demands and technology advancements. Significantly, O-RAN facilitates the utilisation of artificial intelligence and machine learning (AI/ML) algorithms, enabling automated network optimization, predictive maintenance, and enhanced security. This approach enhances network management by reducing the need for manual intervention and associated labour costs. Despite the great potential of the standardized O-RAN architecture \cite{Polese_et_al:2023:UOA}, resistance to deploying O-RAN-based solutions persists due to the following challenges:

\begin{itemize}
    \item The O-RAN architecture and its use cases are relatively new, prompting hesitation from MNOs to invest in O-RAN solutions. Limited real-world deployment experiences have hindered a comprehensive understanding of the full benefits of the O-RAN architecture.
    \item Concerns about security threats arise from the openness of interfaces in the O-RAN architecture. While the O-RAN Alliance is actively working on standardising security aspects, consumers are still awaiting further maturity in O-RAN security.
\end{itemize}

\subsection{Centralised Intelligent Control in O-RAN}

There are three key aspects in O-RAN specifications: i) the BS is logically and functionally split into a central unit (CU), a distributed unit (DU) and a radio unit (RU) based on the 3GPP 7.2 split with additional features, which is known as 7.2x split in O-RAN. The CU and DU are logical nodes that can be implemented by virtual network functions (VNFs) hosted by a cloud while the RU is a physical node that is usually implemented by a physical network function (PNF) comprising vendor-specific hardware and software components; ii) the introduction of the near-real-time (near-RT) and non-RT RAN intelligent controllers (RICs); and iii) the standardisation of a virtualisation platform for the RAN. 

The RICs connect to the CU and DU through open interfaces for data collection and intelligent control. AI/ML integration into O-RAN occurs through the standardised central controllers (non-RT and near-RT RICs). The Service Management and Orchestration (SMO) is an automation framework that supports RAN operations via standardized interfaces such as O1, A1, and the open fronthaul M-Plane. It gathers and enriches data from O-RAN logical elements (e.g., CUs, DUs, and RUs) or external application servers. This data is used to instruct an rApp in the non-RT RIC to train an ML model with specific goals, such as minimizing power consumption in a particular area of the radio network. The model parameters are then sent to an xApp in the near-RT RIC to derive policies for O-RAN elements using near real-time data from E2 nodes. This method is advantageous for various use cases as it accommodates the time-intensive ML model training at the non-RT RIC, followed by policy inference at the near-RT RIC within stricter time constraints. However, uncertainties persist. For example, decisions like shutting off RUs to conserve energy require comprehensive anomaly checks instead of just relying on brief periods of low traffic. Similarly, decisions to add or upgrade DUs and RUs in response to increased traffic entail complex what-if scenarios. Validating large-scale models that involve intricate resource usage poses significant challenges. However, merely implementing add-on xApps and rApps with conventional algorithms may not fully overcome these challenges. Integrating them with AI technology can significantly enhance network performance. Examples include:

\paragraph{Network Energy Saving}
The introduction of advanced technologies like massive multiple-input multiple-output (mMIMO) antennas and network densification enhances network performance but also increases the carbon footprint due to greater computing demands. To mitigate this, energy-saving strategies include: 1) reducing mMIMO antenna activity during low traffic periods by turning off specific RF circuits; 2) selectively deactivating RUs and reorganising cells for optimal coverage; 3) implementing sleep-active patterns across multiple sites while ensuring Quality of Service (QoS); and 4) lowering CPU cycles of additional RAN elements (e.g., DUs and CUs) during off-peak times. AI/ML is important for decision-making in these energy-saving approaches \cite{Lopez-Perez_et_al:2022:NE}.

\paragraph{mMIMO Beamforming Optimisation}

While fully digital mMIMO may be prohibitively expensive for some sites, hybrid antenna arrays, as an alternative, use the extensive degrees of freedom of mMIMO systems but control antennas in groups instead of individually. This approach allows mobile terminals within a cell to be served by a grid-of-beams, with semi-static beams designed to cover specific geographical regions. Since mobile terminals are not uniformly distributed, a grid-of-beams can be iteratively optimised with AI/ML assistance, adapting to changes in user distribution and continuously refining beam configurations based on AI/ML model feedback.

\paragraph{Zero Touch Network Management}

In simple network scenarios, manual configuration by skilled operators can effectively address failures or underperformance. However, the increasing complexity of modern networks often makes manual management impractical, driving the need for zero-touch network management using AI/ML. This approach is essential as AI/ML-optimized configuration can reduce human errors \cite{Asharaf:2022:ZTN}.
Developing AI/ML solutions to address these challenges is not straightforward.  
The integration of AI/ML is targeted in non-RT RIC's rApp and near-RT RIC's xApp for optimal RAN control to achieve specific objectives \cite{Polese_et_al:2023:CRD}.

\section{Concept: Digital Twin for O-RAN}

We explore two distinct yet promising conceptual DT directions: The first focuses on creating a holistic digital replica of the
entire RAN with a hierarchical structure while the second adopts a modular architecture, treating DT as flexible building blocks for constructing and evolving the RAN.

\subsection{Digital Twin Network for the Whole RAN}
\begin{figure}%[t]
\centering
    \hspace*{-1.2cm}
    \includegraphics[width=0.5\textwidth]{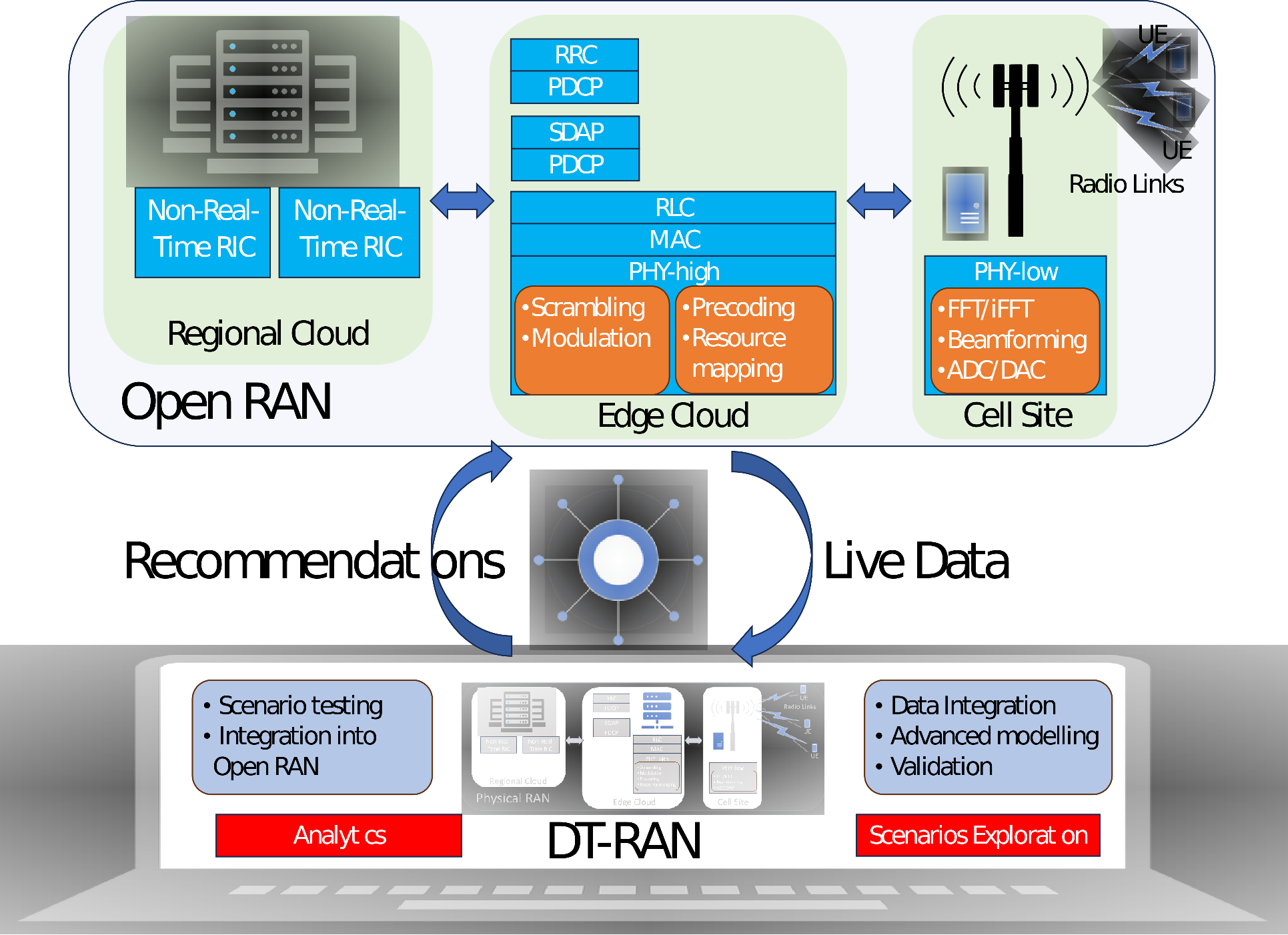}
\caption{\label{DT_Whole_RAN} Digital Twin Network for the whole RAN.}
\end{figure}
\begin{figure}[t]
\centering
    \hspace*{-0.2cm}\includegraphics[width=.51\textwidth]{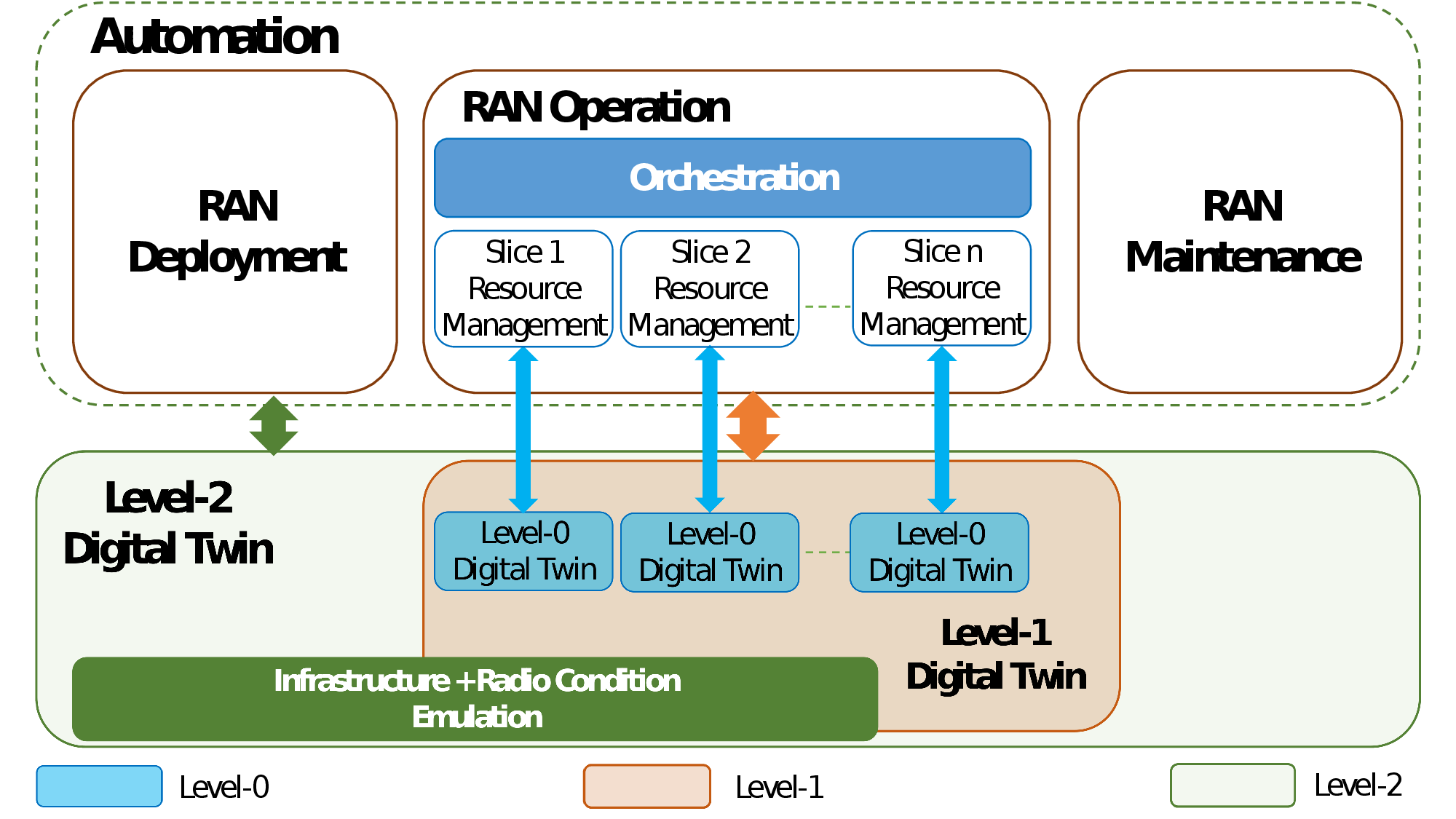}
\caption{\label{DT_ORAN04} Proposed hierarchical DT architecture for O-RAN.}
\end{figure}

In this approach, a mirror of the RAN, which is referred to as a RAN Digital Twin Network (DT-RAN) as illustrated in Fig.~\ref{DT_Whole_RAN}, includes all the logical elements of a RAN. The presence of DT-RAN enables us to i) explore various what-if scenarios and ii) harness intelligent RAN control with minimal human intervention \cite{Lin_et_al:2023:6GDT}. This approach should work well for the case when the creation of a roll-out script for a network on a large scale is required, even initiated initially with a small set of sites only. This process builds confidence in deploying hardware, VNFs in O-Cloud, and applications in RICs. The DT-RAN should exist side-by-side with the RAN and may use new interface(s) for data and information exchange if necessary. Creating a complete replica of the whole RAN is a complex process so we propose that it is managed in progressive stages within a hierarchical structure, each represented by different levels (e.g., Level-0, Level-1, etc) as depicted in Fig. \ref{DT_ORAN04}. Each level corresponds to a specific fidelity requirement of the DT.

\paragraph{Fidelity and Hierarchical Architecture of DT-RAN}

`Fidelity' broadly defines the accuracy and detail level of replication concerning RAN functionalities and services, as well as the components of the RAN that the DT can mimic. Similar to many other systems, the evolution of a DT-RAN follows a life cycle in which not all components of DT-RAN can be present from the initial stages. Consequently, the architecture of DT-RAN should facilitate its progression from a basic form to a sophisticated ecosystem, and the concept of hierarchy can support this evolution. The fidelity correlates with these proposed hierarchical levels. Fig.~\ref{DT_ORAN04} shows that Level-0 DT, which has lowest fidelity, represents basic functions like data enrichment, traffic flow modelling and resource allocation in a RAN slice. A higher fidelity DT at Level-1 introduces greater complexity, managing the entire operation and orchestration of RAN slices under certain resource constraints, adding further functions such as load balancing and network security. Level-2 DT expands further to the whole range of RAN functionalities, mimicking a fully operational DT-RAN, thus has the highest fidelity.
Examples of such Level-2 DT's functionalities include prediction and optimisation (e.g., automated RAN deployment, planning and cell reorganisation, and energy efficiency) and maintenance (e.g., network predictive maintenance, interference management and fault/anomaly detection). This Level-2 DT should closely emulate the performance of a fully operational DT-RAN.
In order to assure the fidelity requirement, design strategies need to be planned meticulously, from data quality, analytics, modelling to validation, matching each hierarchical level.

\paragraph{Components and Operation of DT-RAN}
The RU, the O-RAN network functions, including CU and DU, and the near-RT RIC in the O-RAN network are digitally replicated to increase automation and efficiency, which can be achieved by testing and analysing different what-if scenarios. The digital mirrors of CU-CP, CU-UP and DU also host the Packet Data Convergence Protocol (PDCP), Service Data Adaptation Protocol (SDAP), Radio Link Control (RLC), Media Access Control (MAC) and physical layer (PHY) control functionalities to replicate the RAN operations.

For a single application, several versions of an xApp can be created in the digitally mirrored near-RT RIC so that the DT-RAN uses them to test against the digital mirrors of the O-RAN elements. The best version is selected to update the xApp entity who works with the real RAN's elements.

In the O-RAN architecture, the SMO platform is specified to provide supports for: i) Fault, Configuration, Accounting, Performance, Security (FCAPS) functions in the O-RAN network via O1 interface; ii) non-RT RIC for RAN optimization together with near-RT RIC via A1 interface; and iii) O-Cloud management, orchestration and workflow management via O2 interface. 
In interaction with the DT-RAN, the RAN management functions in the SMO can also provide FCAPS services simultaneously to the physical RAN components and their mirrored counterpart. The functions in the SMO framework can be realised with DT-RAN for model training, inference and updates needed for rApps within the non-RT RIC.

\paragraph{Auxiliary Models of DT-RAN}

Additional models, not necessarily mimicking components of a RAN, can enhance the functionalities of DT-RAN. These include i) an RF propagation model, ii) a user-data traffic model, and iii) a cost model for the RAN's NFs. Each of these models plays a crucial role in the operations of a DT-RAN. For example, the RF propagation model can predict whether the deployment of a new remote radio head (RRH) at a location would lead to an expansion of network coverage. On another front, a user-data traffic model aids in predicting network density and the level of advancing mMIMO in a given area. The enhancements in coverage and throughput can then be justified against the additional cost predicted by the cost model.

\paragraph{Interaction between DT-RAN and Physical RAN}

The interaction between the DT-RAN and the physical RAN facilitates the functioning and evolution of the latter. However, the O-RAN specifications solely define the interfaces between O-RAN logical elements (including O-RU, O-DU, O-CU-CP, O-CU-UP, and near-RT RIC), with the exception of SMO, which interfaces with external systems to acquire enrichment data. This limitation in the O-RAN architecture implies that there might be need for additional solutions to support the concept of housing the DT as building blocks of the RAN, and these possibilities will be explored in greater detail in the following section.

\begin{figure}[t]
\centering
    \hspace*{0cm}\includegraphics[width=.5\textwidth]{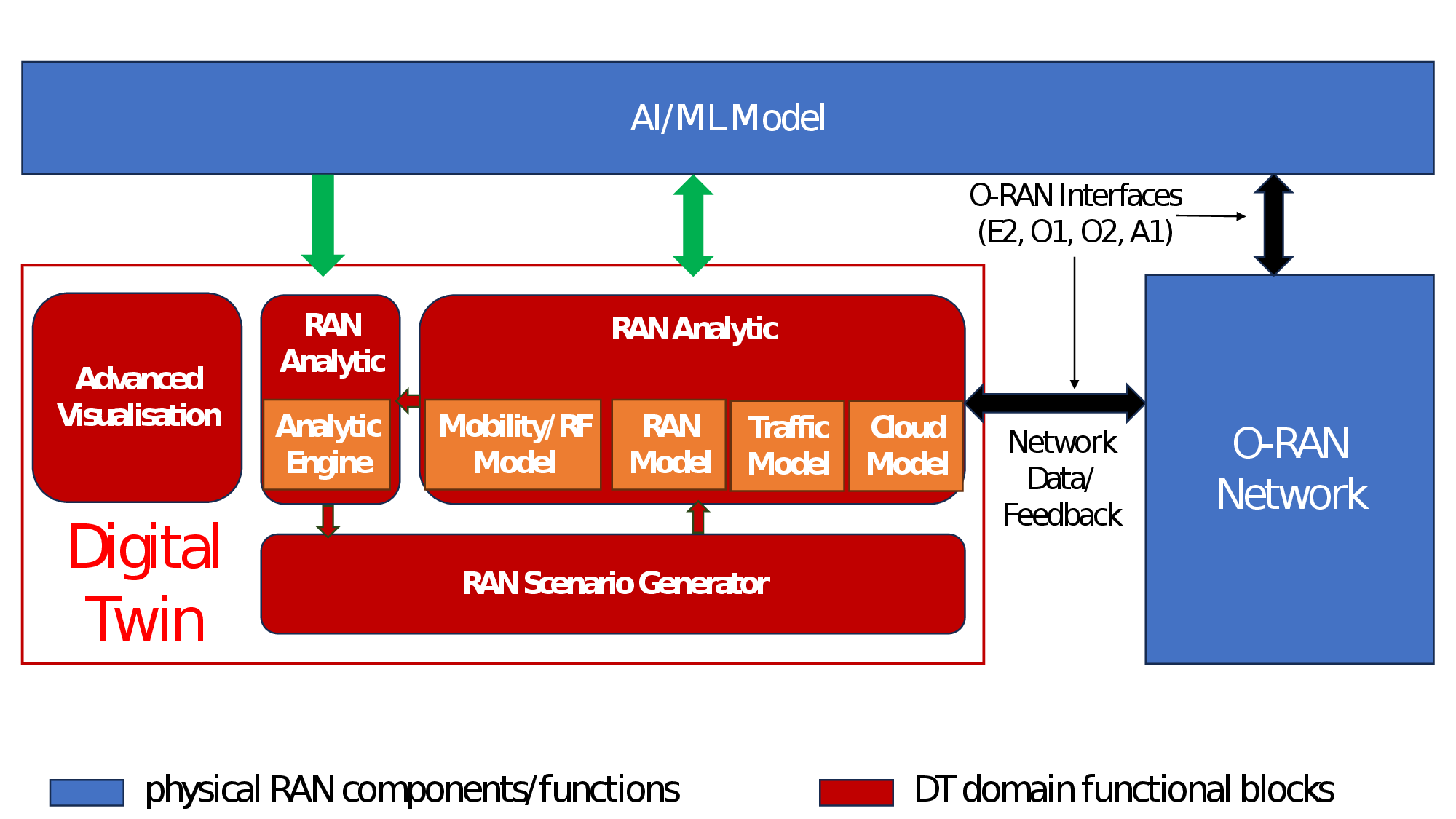}
\caption{\label{DT_ORAN02} DTN building blocks 
}
\end{figure}

\subsection{Digital Twins as RAN Building Blocks}
Developing the DT concept for the entire RAN can be complex. Nonetheless, we can suggest the idea of formulating the DT as building blocks (or modules), employed to construct the DT-RAN eventually, and design it to be open for extension as a system-of-systems, as depicted in Fig.~\ref{DT_ORAN02}.

This approach views the DT as modular building blocks within the RAN that are utilised to work with other RAN components. Existing RAN interfaces are used for the DT. These DTs are designed to be interoperable and open for extension, enabling the gradual assembly of a complete ecosystem (or library) of various DT building blocks. This approach emphasises scalability, flexibility, and modularity, treating the RAN as a system-of-systems that can evolve over time to accommodate new functionalities, technologies, and requirements.

The building blocks can represent individual RAN components or be designed as essential DT elements (similar to plugins), which may not directly correspond to actual RAN components. Below are some examples of the DT building blocks:
\begin{itemize}
    \item {\it Modelling entities}: These entities create precise digital replicas of various aspects of a RAN network. This includes a mobility/RF model for channels between devices, a traffic model for data demand and their QoS constraints, a RAN model representing entities in the O-RAN network, and a cloud model for computing resources. The models stay synchronised with the live physical network through data captured from O-RAN standardised interfaces like O1, O2, E2, and A1 (no new interfaces required).
    \item {\it RAN scenario generator}: Empowered by AI/ML technology, this generator automatically parameterises the models to produce billions of training scenarios. These scenarios challenge AI/ML models under training. The RAN scenario generator can autonomously evolve based on performance feedback from the \textit{RAN analytic module}, generating increasingly challenging training datasets. This process contributes to the continuous improvement of AI/ML intelligence and performance.
    \item {\it Advanced visualisation}: This component is responsible for visualising the data needed by the network operators to facilitate performance monitoring and system diagnostics.
\end{itemize}
Examples of the modelling entities can include (but not limited to):

\paragraph{Modelling Physical RF Propagation}
For realistically modelling the physical RF propagation characteristics in the mobility/RF model, ray tracing has been identified as one of the most promising candidate technologies that can be leveraged for DT. Ray tracing estimates RF propagation characteristics by calculating path gains through a geometrical region with varying velocity, absorption characteristics, and multiple reflecting surfaces. The accuracy of the calculation relies on the ray tracing algorithm itself as well as the measurement data that calibrate the penetration loss, reflection and scattering characteristics of the surfaces of obstacles in the 3-D environment geometry (e.g., buildings). There is also more advanced ray tracing technology that can automatically generate a propagation model with less computational complexity and better accuracy based on AI/ML that is trained with large amount of real measurement data~\cite{Ju-Hyung_Globecom_2023}. 

\paragraph{Modelling RAN and Cloud}
Modeling the dynamic behaviours of O-RAN elements and cloud hardware resource allocations realistically, while maintaining synchronisation with real-time states of the physical network, poses a challenge for RAN and cloud models. A promising approach is to virtualise the network and UE functions, simplifying them to retain only the network and UE state, call flow, and KPI prediction capabilities. Relaxing the real-time constraints of the virtualised mirror of the RAN elements and the cloud resource to near real-time can help reduce the computing complexity and resources of the DT. Data plays a crucial role in syncing the RAN and cloud models with the physical environment.

To address the high computing complexity of the DT, selecting an appropriate computing technology is essential. A GPU-based cloud platform is a feasible choice due to its programmability and virtualisability, capable of handling extensive parallel computing tasks with significantly higher throughput compared to a CPU. The relaxed real-time and latency constraints on the DT, if possible, enable higher degrees of parallelization and GPU computing acceleration, making them more achievable than with physical network elements.

\section{Solution Examples: Integrating Digital Twin into O-RAN}

We explore three solutions for integrating DT concept, utilising the established standardised O-RAN interfaces and leveraging the advantages of openness.

\subsection{Digital Twin in Non-RT RIC}
\begin{figure}[t]
\centering
    \includegraphics[width=.5\textwidth]{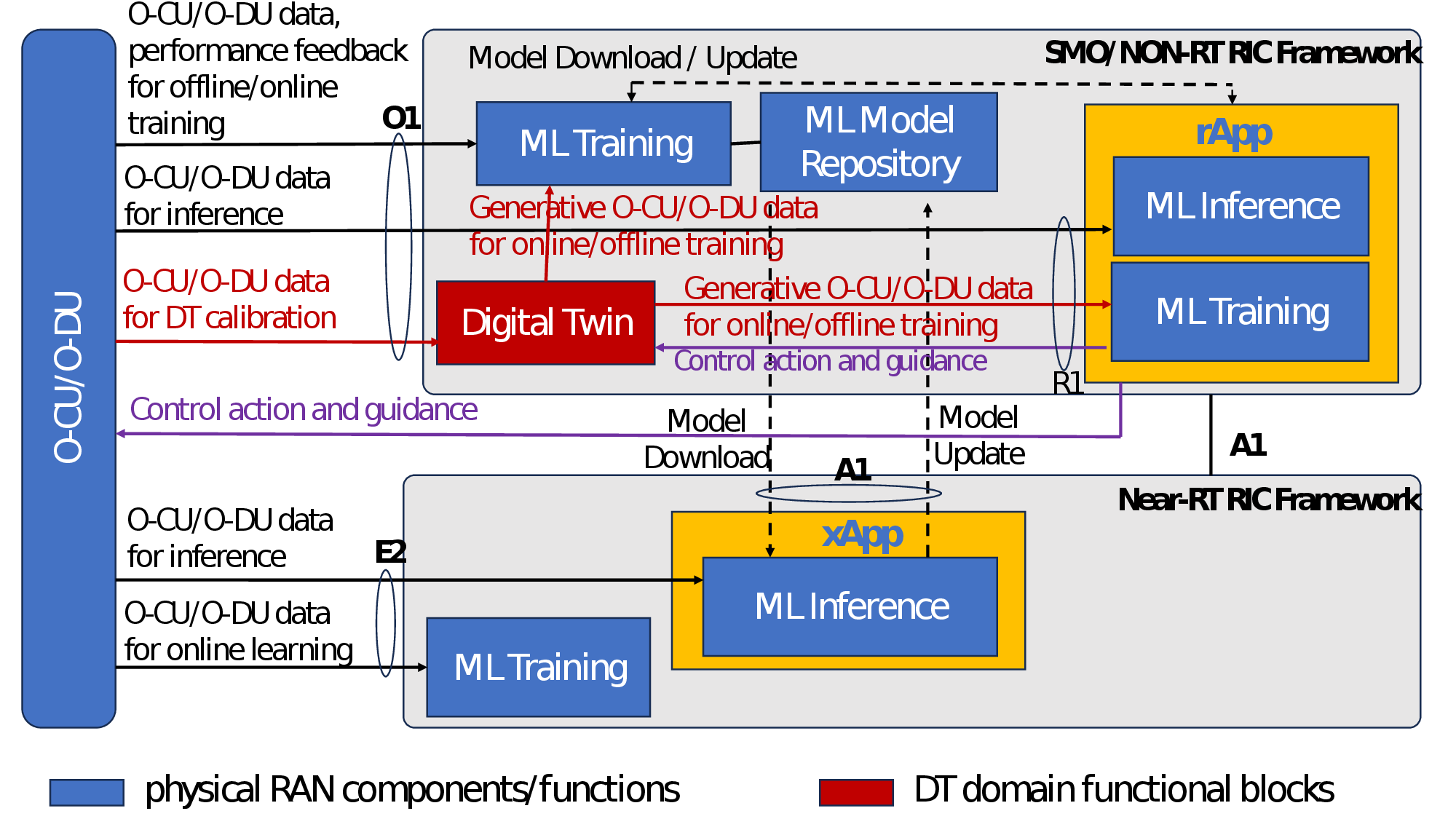}
\caption{\label{DT_ORAN09} DT integrated in non-RT RIC.}
\end{figure}

Figure~\ref{DT_ORAN09} shows the DT residing in the non-RT RIC framework to improve the performance of the training and testing process. The DT, implemented with the aid of advanced computer simulation and modelling technology, can potentially represent an exact digital replica of physical O-RAN network.

The rApps can engage with the DT function within the non-RT RIC framework through the R1 and A1 interfaces, facilitating AI/ML workflow-related services such as model training, testing, and real-time performance assurance. Specifically, rApps interact with the DT in the following ways:

\begin{itemize}
    \item[1) ] \textit{directly accessing training data} generated by DT via the R1 interface;
    \item[2) ] \textit{loading the AI/ML models} into the training host located in the non-RT RIC framework for platform layer model training and testing; and \textit{specifying}, through the R1 interface, whether and when the physical network data or DT data should be used; and
    \item[3) ] \textit{conveying the control actions and policies} to the DT during AI/ML inference; and \textit{predicting the performance impact} before forwarding them to the physical network for real-time performance prediction and assurance.
\end{itemize}

\subsection{Digital Twin in Near-RT RIC}
\begin{figure}[t]
\centering
    \includegraphics[width=.5\textwidth]{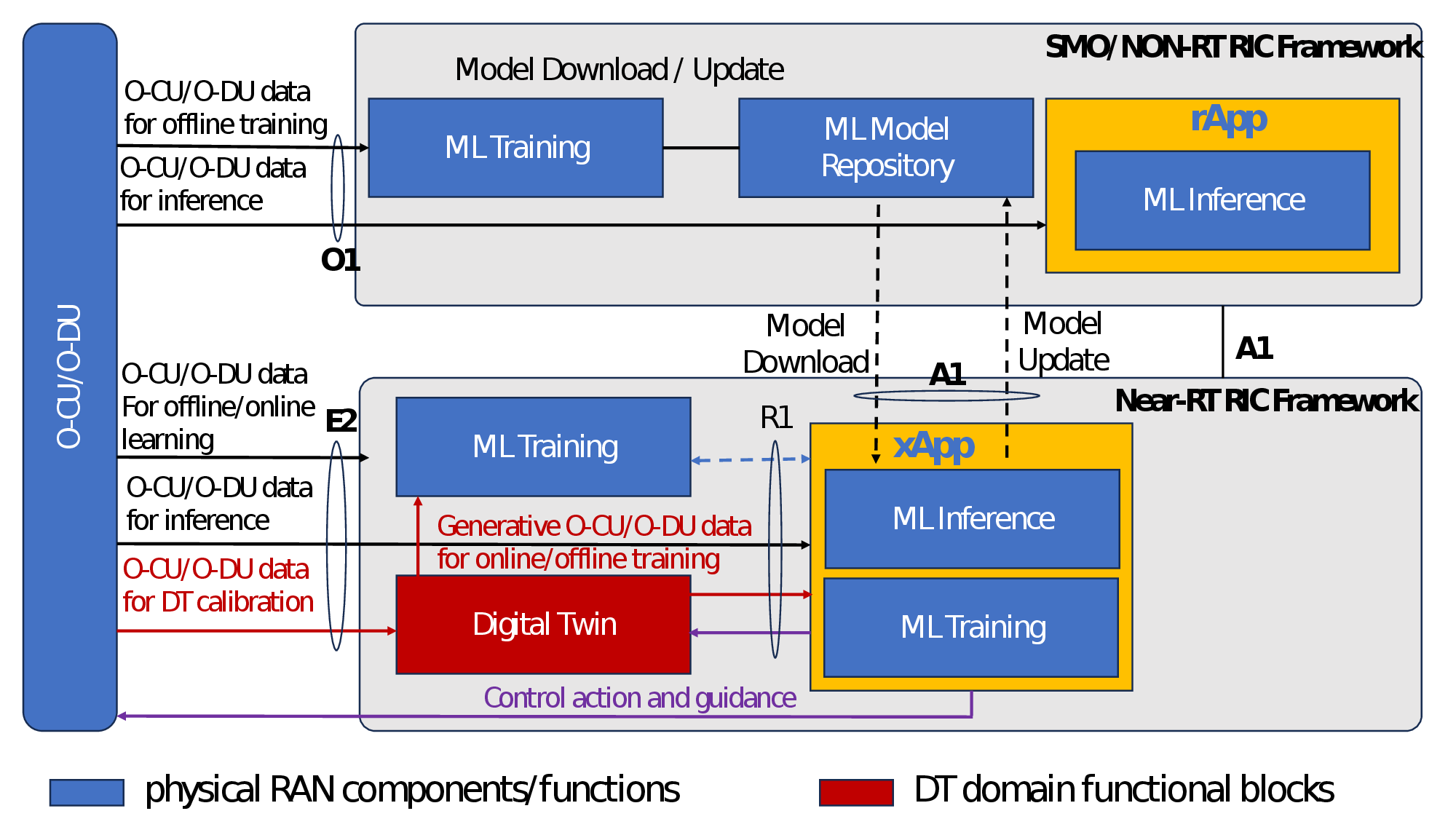}
\caption{\label{DT_ORAN10} DT integrated in near-RT RIC.}
\end{figure}

Figure~\ref{DT_ORAN10} presents another solution featuring the DT located at the Near-RT RIC. In this solution, the E2 interface is used to collect real network data for training (online or offline) and then testing the AI/ML models used by the xApps.

The models in the DT is calibrated with the data captured from the E2 interface to ensure the DT functions are behaving as closely as possible to the real network deployment for reliable RIC AI/ML model training and testing.

Similar to the rApps in previous subsection, the xApps can access the DT functions in the Near-RT RIC framework via the Near-RT RIC API for AI/ML model training and testing. Apart from using the Near-RT RIC API, the same data flow and methods for AI/ML training, testing, performance assurance, and DT calibration introduced in the previous solution still apply.

Note that the DT can also be useful for detecting and mitigating conflicts among xApps. By running scenarios under various network conditions, DT can predict individual and collective xApp behaviours, helping to understand potential conflicts and their impact on network performance and stability. Additionally, DT can assist in developing coordination mechanisms and policies for xApps coming from different vendors, who can also test their xApps on the DT platform before deployment.

\subsection{Digital Twin Outside Non-RT RIC and Near-RT RIC}

\begin{figure}[t]
\centering
    \includegraphics[width=.5\textwidth]{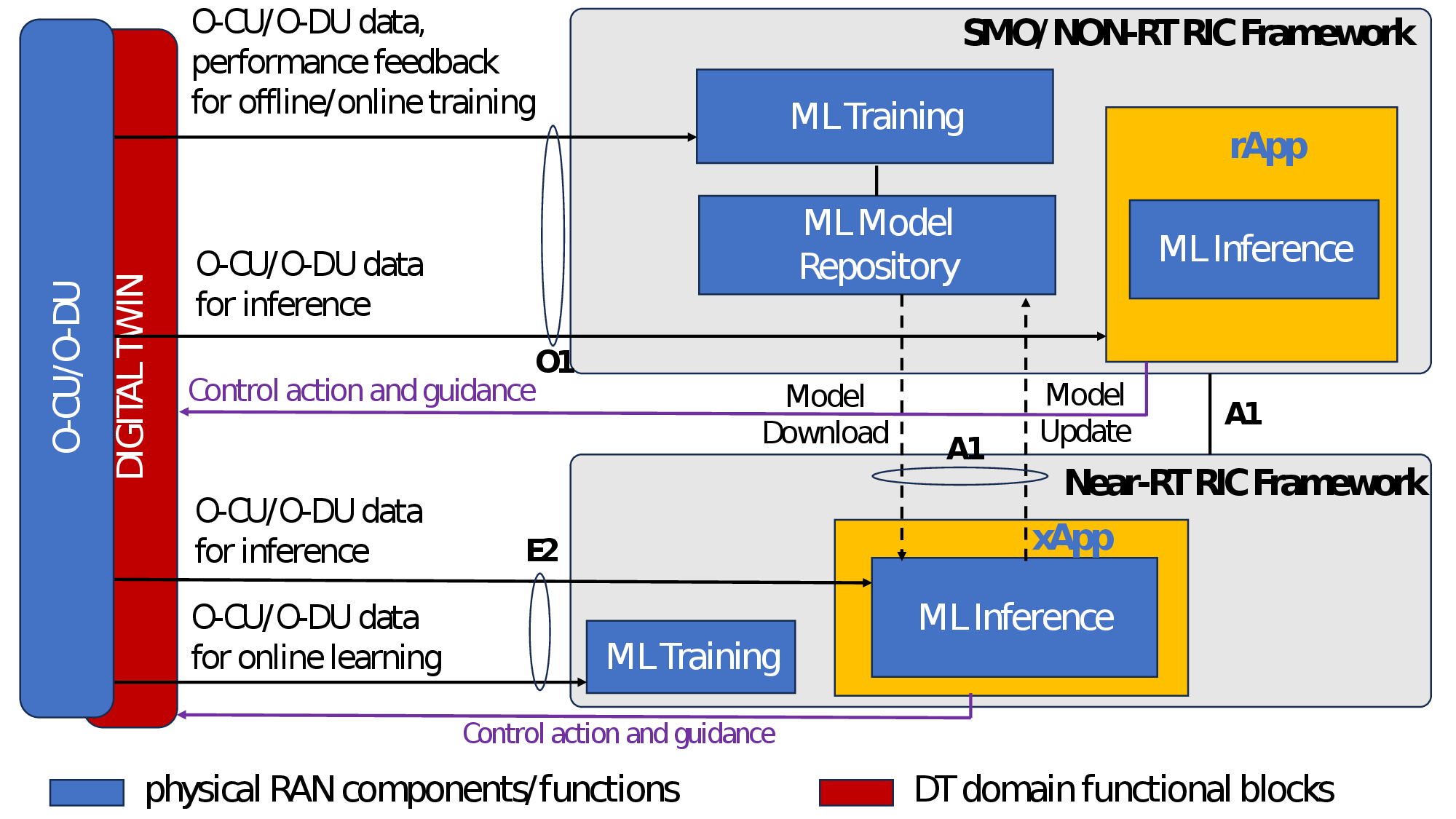}
\caption{\label{DT_ORAN11} DT integrated outside Non-RT RIC and Near-RT RIC.}
\end{figure}

Another viable solution, shown in Fig.~\ref{DT_ORAN11}, involves housing the DT external to the near-RT and non-RT RIC frameworks. In this scenario, the DT, simulating the physical network environment, operates concurrently with the real network. The DT can be hosted independently on a platform, distinct from the O-Cloud, where interactions with the physical RAN occur via the RAN's SMO interface.

From the RIC perspective, there is no distinction between the physical network and the simulated network. RICs communicate with both the DT and physical RAN through standard O-RAN interfaces (O-FH, O1, O2, and E2). While solutions based on the two approaches above follow the interfaces specified by O-RAN Alliance, this standalone approach allows more flexibility in choosing the technologies and frameworks that best suit the DT's needs without being constrained by the RIC's architecture. However, this separation may result in challenges related to ensuring that the external DT has access to all necessary data from the RIC while maintaining strict data privacy and security protocols. External placement can also complicate the synchronisation processes required to keep the DT updated with the latest network changes, thus any time-critical tasks/responsibilities may not be desirable in this approach, unless a new solution/framework can be found to manage the DT. In addition, lack of standardisation may pose challenges for future scalability.

It is also crucial to consider potential communication overhead on external O-RAN Alliance interfaces (O-FH, O1, O2, and E2) during AI/ML model training, where numerous training scenarios are generated. The DT obviously needs robust and secure communication channels to interact with the RIC. As a result, additional latency may be introduced due to the physical and logical separation between the DT and the RIC. Despite this, the same data flow and methods for AI/ML training, testing, performance assurance, and DT calibration, as introduced in the previous solutions, remain applicable.

\section{Other Open Challenges for DT of the RAN}

\subsection{Data Exchange}
To maximise the benefits of DT in O-RAN, efficient data exchange between the Physical RAN and its DT is crucial, especially with external DT placement. Outdated data can degrade performance and cause operational mismatches. Depending on the DT approach, either existing interfaces might be used, or new ones may need to be developed. Additionally, dynamically coordinating the data exchange between the DT and Physical RAN is essential, as different network scenarios require varied data amounts due to strict latency, privacy, and security needs, necessitating adaptive coordination for efficient resource use.

\subsection{Computational Capacity}
While the DT mirrors its Physical RAN, it requires superior computational capacity to forecast and recommend/control Physical RAN operations. Determining the extent of this superiority (thus extra costs) depends on various factors, including the computational capacity of the Physical RAN and the rate of information exchange between the Physical RAN and its DT. This poses an open problem that requires further investigation.
\subsection{The Ecosystem of DTs within Physical RAN}
The hierarchical DT model offers the flexibility and scalability in deployment. Having multiple DTs of different O-RAN components poses several challenges. First, if new interfaces are introduced for the case of external DT placement, there may be need for standardisation of interfaces. Second, for the case of internal placement, the DT can consume the computational resources of the other RAN components, which may impact the critical tasks that requires near-real-time decision making. Third, appropriate protocols/procedures are needed to avoid conflicts on resource utilisation and control. Such protocols/procedures will also help optimise the data exchanges between DTs and their physical O-RAN components. 
\subsection{Security}
The O-RAN architecture presents security risks from application to hardware levels, and the integration of DT introduces additional vulnerabilities due to increased data circulation. Protecting the exchanged data between DT and Physical RAN from cyber threats is crucial. Addressing this challenge involves researching xApp/rApp for threat identification and elimination, devising methods to enhance secure data exchange, and preventing unauthorized O-RAN elements from exploiting network access mechanisms.
\subsection{Physical RAN and its DT Dependency}
It is essential to study the dependency of the Physical RAN on its DT. Factors such as the optimal frequency and level of interaction between DT and Physical RAN, considering constraints on data exchange bandwidth and the adopted hierarchical DT model, need examination. The DT must also anticipate and prepare for the worst-case scenario in the event of system failure. Maintaining the Physical RAN's superiority compared to a standard O-RAN, one that lacks an integrated DT, remains an open and challenging problem.
\section{Conclusion}
The rapid advancement of DT technology is undeniable, driven by both extensive research attention and its broad applicability. The symbiotic relationship between DTs and emerging 6G networks is evident, with a DT network becoming essential for cost-effective deployment, efficient automated operation, and straightforward maintenance of a 5G/6G network. This paper studied two conceptual directions for DT integration into O-RAN, along with three solution examples and five open challenges, laying the groundwork for potential future research in this dynamic landscape.
\bibliographystyle{IEEEtran}
\bibliography{references.bib}
\vskip -2\baselineskip plus -1fil 
\begin{IEEEbiographynophoto}{H. X. Nguyen}
is a Professor of Digital Communication Engineering, Director of the London Digital Twin Research Centre and Head of the 5G \& IoT Research Group at Middlesex Univ., London, UK. He is also a Visiting Professor at International School, Vietnam National University, Hanoi, Vietnam. He received his Ph.D. degree from the Uni. of New South Wales, Australia, in 2007. He leads research activities in digital twinning, 5G/6G/IoT systems, and digital transformation solutions.
\end{IEEEbiographynophoto}
\vskip -2\baselineskip plus -1fil 
\begin{IEEEbiographynophoto}{K. Sun} 
serves as the Principal Wireless Architect at Rakuten Symphony, Rakuten Group, designing architectures for SMO, Non-RT RIC, and AI/ML products/solutions. He is a delegate in the O-RAN Alliance, focusing on AI/ML, SMO, RIC, and digital twin specifications, and co-chairs the O-RAN nGRG RS1, leading next-gen DT-based O-RAN technology research. Sun earned his M.Sc. in Wireless Communications from University College London in 2011.
\end{IEEEbiographynophoto}
\vskip -2\baselineskip plus -1fil 
\begin{IEEEbiographynophoto}{D. To} received his Ph.D. degree in Advanced Telecommunications from Swansea University, U.K., in 2011. Since then, he has been taking various roles in different companies in the mobile communications industry. Dr. To is is currently taking the role as a Principal System Architect at Baseband and Call Processing department, CTO Office, Rakuten Symphony, Rakuten Group, responsible for architectural and algorithmic designs for 5G RAN networks.
\end{IEEEbiographynophoto}
\vskip -2\baselineskip plus -1fil 
\begin{IEEEbiographynophoto}{Q.-T. Vien} received his Ph.D. degree in
Telecommunications from Glasgow Caledonian University in 2012 and is a Senior Lecturer at Middlesex University. He authored a textbook, four books, and over 100 research papers. His research focuses on physical-layer security, network coding, and other advanced telecommunications areas. Dr. Vien received the Best Paper Award at the IEEE/IFIP Conference in 2016 and was named Exemplary Reviewer by IEEE Communications Letters in 2017.
\end{IEEEbiographynophoto}
\vskip -2\baselineskip plus -1fil 
\begin{IEEEbiographynophoto}{T. A. Le}
received his Ph.D. in Telecommunications from King's College London in 2012 and was a Post-Doctoral Research Fellow at the University of Leeds. He is now a Senior Lecturer at Middlesex University. His research spans integrated sensing and communication, energy harvesting, physical-layer security, and ML for wireless communications. Dr. Le was honored as an Exemplary Reviewer by IEEE Communications Letters in 2019/2023 and serves as an Associate Editor of the IEEE Wireless Communications Letters.
\end{IEEEbiographynophoto}
\vfill
\end{document}